\documentclass[twocolumn,aps,amsmath,amssymb,prl,superscriptaddress]{revtex4}
\usepackage{graphicx}
\usepackage{amsmath}
\usepackage{bm}

\begin{document}

\title{Gate-controlled spin-orbit interaction in a parabolic GaAs/AlGaAs quantum well}

\author{M. Studer\footnote{mattstud@phys.ethz.ch}}
\affiliation{IBM Research, Zurich Research Laboratory,
S\"aumerstrasse 4, 8803 R\"uschlikon, Switzerland}
\affiliation{Solid State Physics Laboratory, ETH Zurich, 8093
Zurich, Switzerland}
\author{G. Salis\footnote{GSA@zurich.ibm.com}}
\affiliation{IBM Research, Zurich Research Laboratory,
S\"aumerstrasse 4, 8803 R\"uschlikon, Switzerland}
\author{K. Ensslin}
\affiliation{Solid State Physics Laboratory, ETH Zurich, 8093
Zurich, Switzerland}
\author{D. C. Driscoll}
\affiliation{Materials Department, University of California, Santa
Barbara, California 93106, USA}
\author{A. C. Gossard}
\affiliation{Materials Department, University of California, Santa
Barbara, California 93106, USA}

\begin{abstract}
We study the tunability of the spin-orbit interaction in a
two-dimensional electron gas with a front and a back gate electrode
by monitoring the spin precession frequency of drifting electrons
using time-resolved Kerr rotation. The Rashba spin splitting can be
tuned by the gate biases, while we find a small Dresselhaus
splitting that depends only weakly on the gating. We determine
absolute values and signs of the two components and show that for
zero Rashba spin splitting the anisotropy of the spin dephasing rate
vanishes.
\end{abstract}

\maketitle

Spin-orbit (SO) interaction is one of the key ingredients for future
spintronic devices. In a two-dimensional electron gas (2DEG), SO
interaction manifests itself as a spin splitting, and two different
asymmetries can be responsible for it: The inversion asymmetry of a
zincblende crystal leads to the so-called Dresselhaus spin
splitting~\cite{Dresselhaus1955}, and an electric field, $E_{QW}$,
along the growth direction enables Rashba-type spin
splitting~\cite{Bychkov1984}. $E_{QW}$ is either generated by an
asymmetrically grown layer structure (e.g., doping profile), or can
be controlled externally by appropriate gating. The latter allows
for a very efficient and scalable approach to control
spins~\cite{Datta1990,Schliemann2003}.

The possibility of tuning the SO interaction in a 2DEG with a front
gate (FG) electrode was first exploited by Nitta et
al.~\cite{Nitta1997}, who studied Shubnikov--de~Haas (SdH)
oscillations. Similar experiments were done in a 2D hole system with
a FG and a back gate (BG) electrode \cite{Papadakis1999}. These
experiments could not differentiate between the Dresselhaus and
Rashba components. By tuning the sheet density of a 2DEG and a
careful analysis of weak antilocalization peaks, Rashba and
Dresselhaus SO interaction were separated in a transport
experiment~\cite{Miller2003}. It has been proposed that such
information could also be obtained from measurements of conductance
anisotropy in quantum wires~\cite{Scheid2008}. Using photocurrents,
it is possible to characterize the sources of SO interaction, but
not to make a quantitative statement about the SO
strength~\cite{Belkov2008,Ganichev2004}. In a 2DEG, the spin
lifetime, which can be measured optically, is limited mainly by the
Dyakonov--Perel (DP) dephasing mechanism~\cite{Dyakonov1972} and
therefore by the strength of the SO interaction. An optical study of
the spin lifetime as a function of gate voltages therefore
indirectly provides information on the tunability of the SO
interaction~\cite{Karimov2003,Larionov2008}.

A more direct method to obtain quantitative access to both the
Rashba and the Dresselhaus SO interaction strength is to measure the
drift-induced effective SO magnetic field (in the following referred
to as SO field)~\cite{Meier2007}. Here, we employ this method to
study the tunability of the SO interaction by means of an external
electric field, $E_{ext}$, perpendicular to the plane of a 2DEG
confined in a parabolic potential. By using FG and BG electrodes,
independent control of both $E_{ext}$ and the carrier sheet
density~\cite{Salis1997} is obtained. Although the electrons are
confined in a parabolic well that does not change its shape with
bias, we find a Rashba SO field that depends linearly on $E_{ext}$.
The Dresselhaus SO field is only weakly affected by the gate bias.
Together with a measurement of the mobility of the 2DEG, we obtain
quantitative values for the Rashba coupling and its sign. Our
optical measurements allow the simultaneous determination of the
different contributions to SO interaction and the spin lifetime. In
agreement with the DP dephasing
mechanism~\cite{Kainz2004,Larionov2008}, the in-plane anisotropy of
the spin lifetime disappears when $\alpha=0$, validating previous
experiments that extracted ratios of SO splittings from measurements
of spin-dephasing.~\cite{Larionov2008}

\begin{figure}
\includegraphics[width=84mm]{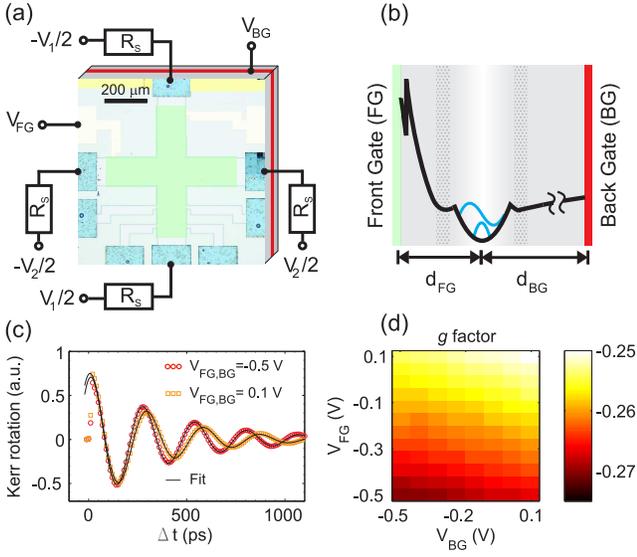}
\caption{\label{fig:fig1} (color online) (a) False-color image of
the sample as seen in an optical microscope: Cross-shaped mesa
structure covered by a FG (green) and Ohmic contacts on each end
(blue). The BG is indicated in red. (b) Schematic conductance-band
profile of the grown layer structure including gate electrodes and
wave functions of the two occupied subbands. (c) TRKR for
$V_{FG}=V_{BG}=-0.5$~V and 0.1~V, respectively. (d) $g$ factor of
the electrons confined in the QW as a function of the gate
voltages.}
\end{figure}

The structure measured is a molecular-beam epitaxy-grown
Al$_{x}$Ga$_{1-x}$As quantum well (QW) with parabolic
confinement~\cite{Salis1997,Salis01}. The QW is 100~nm thick, and
the Al concentration varies from $x=0$ in the center of the well
(located $d_{FG}=105$~nm below the surface) to $x=0.4$ at the edges.
The QW is modulation doped with Si on both sides. A 490~nm thick
layer of low-temperature-grown GaAs isolates the QW from the BG,
which consists of a highly $n$-doped GaAs layer and is located
$d_{BG}=1100$~nm below the QW, see Fig. {\ref{fig:fig1}}(b). Using
photolithography and wet etching, a cross-shaped mesa structure [see
Fig.~\ref{fig:fig1}(a)] with standard AuGe Ohmic contacts to the
2DEG and the BG is defined. A semi-transparent FG consisting of a
2~nm Ti adhesion layer and 6~nm of Au covers the cross. We
characterize the electronic properties of the sample at 30~K in a
magnetic field perpendicular to the sample plane, using the lower
arm of the cross as a Hall bar. From measurements in the dark and
under illumination, we find a persistent photoconduction that
increases the carrier sheet density in the QW by a factor of about
two and lowers the effectiveness of the gates under illuminated
conditions. Both effects are attributed mainly to the ionization of
DX centers in the doping layers. After illumination, the sheet
density is tunable between $7.1 \times 10^{11}$ cm$^{-2}$ and $7.5
\times 10^{11}$ cm$^{-2}$ by applying bias voltages $V_{FG}$ between
the FG and the 2DEG as well as $V_{BG}$ between the BG and the 2DEG.
At 2.4~K, SdH oscillations in $1/B$ exhibit a beating that
corresponds to two frequencies, indicating that two subbands are
occupied~\cite{Salis1997}. The sum of its sheet densities, extracted
from the SdH oscillations, equals the Hall density. From this, we
conclude that a possible parallel conductivity layer has a much
lower mobility than the 2DEG. In the experiments presented in the
following, the 2DEG is in the illuminated state.

We use time-resolved Kerr rotation (TRKR) to probe the spin dynamics
of the carriers confined in the QW \cite{Meier2007}. The pump pulses
(average power 700~$\mu$W; repetition rate 80~MHz) are focused on a
30-$\mu$m-wide spot in the center of the cross. The probe beam
(70~$\mu$W) is focused onto the same spot. The TRKR signal is well
described by an exponentially decaying cosine $A \exp(-\Delta t/
T_2^*)\cos(g \mu_B B_{\textrm{tot}} \Delta t /\hbar)$, where $A$ is
the amplitude of the Kerr signal, $T_2^*$ the ensemble spin
lifetime, $g$ the electron-spin $g$ factor, $\mu_B$ the Bohr
magneton, and $B_{tot}$ is the total magnetic field that is composed
of an external magnetic field, $B_{ext}$, and a SO
field~\cite{Meier2007}. The measurement of the SO interaction relies
on the fact that the SO field and thus $B_{tot}$ depend on the
direction and magnitude of the electron drift, induced by applied
voltages $V_1$=$V_0 \cos(\varphi)$ and $V_2$=$V_0 \sin(\varphi)$
symmetrically to the four arms of the cross via serial resistors
$R_{S}=4.7$~k$\Omega$, see Fig.~\ref{fig:fig1}(a). This creates a
well-controlled in-plane electric field $\mathbf{E}_D(\varphi,V_0)$
in the center of the cross \cite{Studer2009}. The application of
$\mathbf{E}_D$ breaks the symmetry in $k$ space and shifts the
electron population in the $s$-th subband by $\delta \mathbf{k}_{s}
=-m^* \mu_s \mathbf{E}_D/ \hbar$, where $\mu_s$ is the mobility of
the subband and $m^*$ the electron effective mass. Because of this,
the average spin of the electrons in this subband is exposed to an
effective magnetic field that can be divided into a Dresselhaus term
$\mathbf{B}_{D,s}$ and a Rashba term $\mathbf{B}_{R,s}$:

\begin{equation}
\label{eq:linRD} \mathbf{B}_{D,s} = \frac{2\beta_s}{g_s \mu_B}
\left(
\begin{array}{c}
\delta k_{y,s} \\
\delta k_{x,s} \\
\end{array} \right)
\quad \mathbf{B}_{R,s} = \frac{2\alpha_s}{g_s \mu_B} \left(
\begin{array}{c}
\quad \delta k_{y,s} \\
-\delta k_{x,s} \\
\end{array} \right)
.
\end{equation}
We use a coordinate system with $x||$[1$\overline{1}$0], $y||$[110]
and $z||$[001], and restrict our discussion to two subbands. For two
occupied subbands there is an additional contribution to the SO
interaction \cite{Calsaverini2008}, that, however, depends only
weakly on $E_{ext}$ and would appear in our measurements as a
constant contribution to the Rashba SO field. Typical inter-subband
scattering times are on the order of ps \cite{Salis1999}, which is
two orders of magnitude faster than the spin precession period in
our experiment. This implies that one precessing electron spin is on
average equally present in both subbands during its lifetime and
that therefore the TRKR signal represents an average of the two
occupied subbands. Only one Larmor frequency is observable, even if
the $g$ factors of the two subbands are different. As pointed out in
Ref.~\onlinecite{Doehrmann2004}, the fast intersubband scattering is
also a source of spin decoherence. Assuming that both subbands
contribute equally to the TRFR signal, we measure a SO field that is
proportional to the average SO field of the two subbands:
$\mathbf{B}_D=(\mathbf{B}_{D,1}+\mathbf{B}_{D,2})/2$ and
$\mathbf{B}_R=(\mathbf{B}_{R,1}+\mathbf{B}_{R,2})/2$. We apply
$\mathbf{B}_{ext}$ in the plane of the 2DEG at an angle $\theta$
with respect to the $x$ axis. If $B_{ext}\gg B_D$ and $B_R$,
$B_{tot}$ can be approximated  by~\cite{Meier2007}
\begin{equation}\label{eq:AB}
\begin{split}
B_{tot}(\theta,\phi)  \approx & B_{ext}+(B_D+ B_R)\cos{\theta}\sin{\varphi} + \\
&(B_D-B_R)\sin{\theta}\cos{\varphi}.\\
\end{split}
\end{equation}
Because of the different symmetries, the drift-induced modification
of $B_{tot}$ is proportional to $B_D+B_R$ for $\theta=0^\circ$, and
to $B_D-B_R$ for $\theta=90^\circ$.

\begin{figure}
\includegraphics[width=84mm]{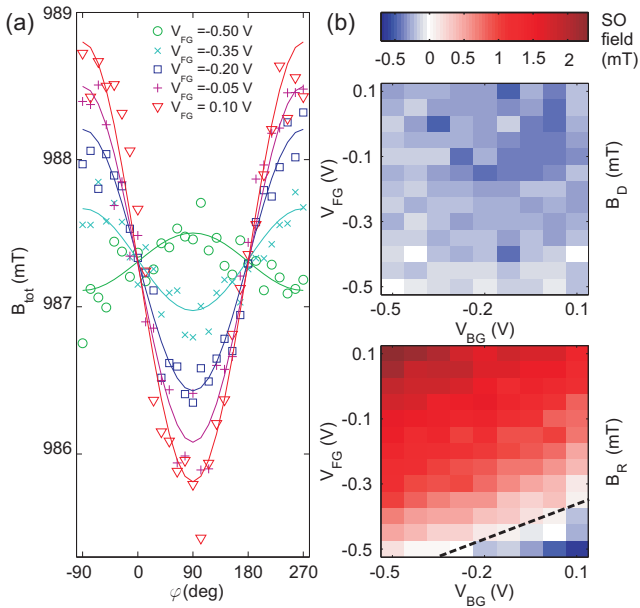}
\caption{\label{fig:fig2} (color online) (a) $B_{tot}$ for
$V_{BG}=-0.375$~V and different $V_{FG}$ biases. The magnetic field
is applied in the $\theta=180^\circ$ direction. The amplitude of the
oscillation in $\varphi$, including its sign, is tunable with the
FG. (b) Experimentally obtained SO fields as a function of $V_{FG}$
and $V_{BG}$. Tuning of the symmetry of the QW allows $B_R$ to be
tuned over a large range, including a sign change (dashed line),
whereas $B_D$ shows only small variations.}
\end{figure}

All the experiments are carried out at 30~K and with
$B_{ext}=0.987$~T. Figure~\ref{fig:fig1}(c) shows the TRKR signal as
a function of the pump-probe delay for two gate configurations
($V_{FG}=V_{BG}=-0.5$ and 0.1~V). The signals are well fit by an
exponentially decaying cosine, and as $B_{ext}$ is known, the $g$
factor can be obtained. Figure \ref{fig:fig1}(d) shows the $g$
factor as a function of the gate voltages. We assume that it is
negative~\cite{Salis01}. The $g$ factor becomes less negative both
for increasing $V_{FG}$ and for increasing $V_{BG}$, indicating that
a larger carrier sheet density and therefore a higher energy of the
electrons are the main cause~\cite{Yang1993}. A lateral displacement
of the subband wave function should also change $g$ and manifests
itself in a dependence where $V_{FG}$ and $V_{BG}$ modify the $g$
factor into opposite directions~\cite{Salis01}. The latter effect
plays a minor role in our doped sample, in contrast to the undoped
samples studied in Ref.~\onlinecite{Salis01}.

Figure \ref{fig:fig2}(a) displays the measured $B_{tot}$ as a
function of the angle $\varphi$ of the direction of $\mathbf{\delta
k}$ for $V_0=1.8$~V, where $E_D\approx$~87~V/m in the center of the
cross. Gating-induced variations of $E_D$ and the mobility result in
a variation in $\delta k$ below 5\%. The geometry of our sample
keeps the density in the center of the cross constant during a
$\varphi$ rotation, preventing a $g$ factor modulation. To test
this, we apply $B_{ext}$ in the opposite direction and find the same
oscillation in $\varphi$ with a sign-reversed amplitude, consistent
with Eq.~(\ref{eq:AB}) (data not shown). The solid lines are fits
using Eq.~(\ref{eq:AB}). As $\theta=180^\circ$, $B_{tot}$ oscillates
in $\varphi$ with an amplitude given by $B_D + B_R$. As seen in
Fig.~\ref{fig:fig2}(a), this amplitude strongly depends on the
$V_{FG}$ applied, suggesting a large variation in the SO
interaction. The SO field does not depend on the magnitude of
B$_{ext}$, and no pump-power dependence of the measured SO fields is
observed~\cite{Studer2009}. The same measurements were done in one
cool-down for a matrix of FG and BG voltages for two samples glued
onto the same chip carrier but with different orientations of the
crystallographic axes such that $\theta=180^\circ$ and
$\theta=90^\circ$, respectively. From these measurements, $B_D$ and
$B_R$ can be obtained separately [See Fig.~\ref{fig:fig2}(b)]. We
observe a large variation in $B_R$, including a sign change (dashed
line). $V_{BG}$ and $V_{FG}$ have the opposite effect on $B_R$,
suggesting that a tuning of the symmetry is responsible for the
variation. Compared to $B_R$, $B_D$ is rather constant. In the
following we will first discuss $B_R$ and then come back to $B_D$.

An electric field $E_{ext}$ applied perpendicularly to the plane of
the QW shifts the potential minimum along the $z$ direction, but
does not change the shape of the parabolic
confinement~\cite{Salis1997}. Nevertheless it is expected that the
Rashba SO coefficient $\alpha$ changes linearly~\cite{Winkler2003},
$\alpha=E_{\textrm{QW}}r_{QW}$. This can be explained by the notion
that the electric field in the valence band determines the Rashba
splitting in the conduction band~\cite{Winkler2003}. Here, $r_{QW}$
is a constant that depends on the material of the QW and
$E_{QW}=E_\textrm{ext}+E_{AP}$ can either be generated by an
asymmetrically grown potential ($E_{AP}$) or by gating the structure
($E_\textrm{ext}$). $E_{ext}$ is related to $V_{BG}$ and $V_{FG}$ by
\begin{equation}
\label{Eext}
E_\textrm{ext}=\frac{V_{BG}}{f_{BG}d_{BG}}-\frac{V_{FG}}{f_{FG}d_{FG}},
\end{equation}
where $f_{BG}$ and $f_{FG}$ are screening factors of the BG and the
FG, respectively.

\begin{figure}
\includegraphics[width=84mm]{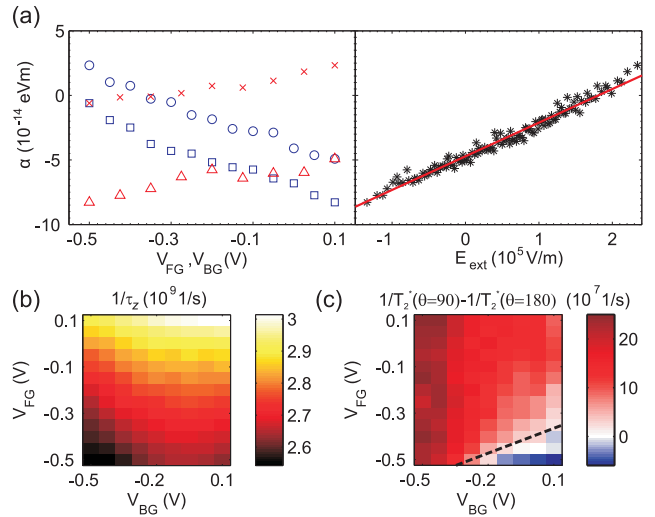}
\caption{\label{fig:fig3}(color online) (a) Left panel: ${\alpha}$
as a function of the FG (BG) voltage with fixed BG (FG) voltage
($\triangle$: $V_{FG}=0.1$ V; $\times$ $V_{FG}=-0.5$~V; $\circ$:
$V_{BG}=0.1$~V; $\square$: $V_{BG}=-0.5$~V). Right Panel: ${\alpha}$
as a function of $E_{ext}$. The red line represents a least-squares
fit as described in the text. (b) Averaged spin-dephasing rate for
$\theta=90^\circ$ and $\theta=180^\circ$ (c) In-plane spin-dephasing
asymmetry. The dashed line marks the same gate voltage regime as in
Fig.~\ref{fig:fig2}(b).}
\end{figure}
As $\delta k_s$ is given by the measured mobility and $E_D$,
${\alpha}=(\alpha_1+\alpha_2)/2$ can be obtained from $B_R$,
assuming that the mobility is the same in both subbands. In the left
panel of Fig.~\ref{fig:fig3}(a), we plot ${\alpha}$ as a function of
$V_{FG(BG)}$, while $V_{BG(FG)}$ is kept fixed. We find parallel
lines for two BG (FG) sweeps, suggesting a linear dependence of
${\alpha}$ on the gate voltages, as predicted by Eq.~(\ref{Eext}).
$E_{ext}$ can be estimated by evaluating the density dependence of
the 2DEG on $V_{FG}$ and $V_{BG}$ and comparing it with a plate
capacitor model~\cite{Papadakis1999}. In the dark we find
$f_{BG}=f_{FG}=1$, i.e. no screening is observed. Under
illumination, $f_{FG}$ is between 15 and 20 and f$_{BG}\approx5$. We
plot all $B_{R}$ data points as a function of $E_{ext}$ by fixing
$f_{BG}=5$ and treating $f_{FG}$ as a fitting parameter to fit all
points to one line [see right panel of Fig.~\ref{fig:fig3}(a)]. This
yields $f_{FG}=21$, which is similar to the value obtained from the
density dependence, indicating that ${\alpha}$ is mainly governed by
$E_{ext}$ and much less so by the density. The sign change of
${\alpha}$ is explained by a change of the symmetry of the QW. From
the slope in Fig.~\ref{fig:fig3}(a), $r_{QW}=25$ e{\AA}$^2$ is
extracted. This is on the same order of magnitude as 5 e{\AA}$^2$,
the value cited for GaAs in Ref.~\onlinecite{Winkler2003}.
${\alpha}$ is positive for $E_{QW}$ pointing from the substrate to
the surface.

In Fig. \ref{fig:fig3}(b) we plot the average of two spin-dephasing
rate measurements $1/ T_2^*$ with $\theta=90^\circ$ and
$\theta=180^\circ$, which is equivalent to the spin-dephasing rate
in the $z$ direction, $1/ \tau_z$. A higher dephasing rate is
observed for higher densities, as expected from the DP
mechanism~\cite{Kainz2004}. In Fig.~\ref{fig:fig3}(c), the
difference of these two measurements is plotted. The in-plane
anisotropy of the spin-dephasing results from the interplay between
the two SO contributions~\cite{Studer2009,Larionov2008}. From
theory, a vanishing anisotropy is expected for
$\alpha=0$~\cite{Kainz2004}. The dashed line in
Fig.~\ref{fig:fig3}(c) indicates bias regions with no Rashba spin
splitting, corresponding also to the region where the in-plane
anisotropy of the spin decay disappears, confirming the theory and
supporting our measurements of $\alpha$.

We now come back to discuss the size of $B_D$, which can be
understood by taking the full Dresselhaus term into consideration.
Starting with the cubic Dresselhaus term \cite{Kainz2004}, we
include the confinement in the $x$-$y$ plane by replacing $k_z^2$ by
the expectation value $\langle k_{s,z}^2 \rangle$ of the $s$-th
quantized subband wave function. Integrating the effective magnetic
field over the shifted Fermi circle using $\delta k_s \ll k_{F,s}$
yields the Dresselhaus SO field including the higher-order terms
\begin{equation}
\label{eq:DriftDH} \mathbf{B}_{D,s} = \frac{2\gamma_s}{g_s \mu_B}
(k_{F,s}^2/4- \langle k_{z,s}^2 \rangle )\left(
\begin{array}{c}
\delta k_{y,s} \\
\delta k_{x,s} \\
\end{array} \right),
\end{equation}
where $\gamma_s$ is a material-dependent parameter~\cite{Kirch2007}
and $k_{F,s}=\sqrt{2 \pi n_{s}}$ is the Fermi wave vector of the
$s$-th subband with the subband density $n_{s}$. Interestingly, two
terms contribute to the SO field: one proportional to $\langle
k_{z,s}^2 \rangle$ and one proportional to $k_{F,s}^2 $. For
sufficient confinement, $\langle k_{s,z}^2 \rangle \gg k_{F,s}^2$,
Eq.~\ref{eq:linRD} is recovered with $\beta_s=-\gamma_s \langle
k_{z,s}^2 \rangle$.

To estimate $B_D$ with Eq.~\ref{eq:DriftDH} we use the SdH densities
at 2.4~K, $n_1=4.2$ and $n_2=2.3\times10^{15}$~m$^{-2}$ and
extrapolate them to the Hall density $7.3\times10^{15}$~m$^{-2}$
measured at 30~K. A numerical simulation of the wave functions in
the QW yields $\langle k_{z,1}^2
\rangle$=2.2$\times10^{15}$~m$^{-2}$ and $\langle k_{z,2}^2
\rangle=7.5\times10^{15}$~m$^{-2}$. With a measured mobility of 8.2
m$^2$/Vs at 30~K, assuming that both subbands have this mobility and
$\gamma_s=10^{-30}$~eVm$^3$, taken from literature \footnote[0]{For
an overview see in the supplementary material of
Ref.~\onlinecite{Kirch2007}}, Eq.~(\ref{eq:DriftDH}) predicts a
value of $-0.4\pm0.5$~mT for $B_D$. This is in good agreement with
the measured $B_D\approx-0.3$~mT at $V_{FG}=V_{BG}=0$~V. The
contributions of the two subbands have opposite sign, leading to
this small value. In Fig. \ref{fig:fig2}(b) $B_D$ tends to slightly
more negative values with positive gating. This trend is explainable
by the higher density and thus higher $k_{F,s}$ and the small
confinement leading to a lower $\langle k_{z,s}^2 \rangle$ because
of the screening of the parabola by the electrons.

To conclude, we have measured the SO field originating from the
Dresselhaus and the Rashba SO interaction in a system where an
electric field $E_{ext}$ perpendicular to the QW plane as well as
the carrier sheet density can be controlled with a FG and a BG. A
small Dresselhaus SO field and a Rashba-induced SO field that
linearly depends on $E_{ext}$ are found. Taking into account the two
occupied subbands, the small values of the Dresselhaus SO field can
be understood qualitatively. We determine the sign of
subband-averaged ${\alpha}$ and show full tunability of ${\alpha}$
through zero. This result is confirmed by a vanishing anisotropy of
the spin-dephasing at ${\alpha}=0$.

We gratefully acknowledge helpful discussions with P. Studerus and
A. Fuhrer, and thank R. Leturcq for experimental help. This work was
supported by the CTI and the SNF.
\bibliographystyle{prsty}

\end{document}